\documentclass[preprint,aps,a4paper]{revtex4-1}
\usepackage{mathrsfs}
\usepackage{graphicx}
\usepackage{lineno}
\begin{document}

\title{Momentum-Resolved  Visualization of Electronic Evolution in Doping a Mott Insulator}

\author{Cheng Hu$^{1,2\dag\ddag}$, Jianfa Zhao$^{1,2\dag}$, Qiang Gao$^{1,2}$, Hongtao Yan$^{1,2}$, Hongtao Rong$^{1,2}$, Jianwei Huang$^{1,2}$, Jing Liu$^{1,2,3}$, Yongqing Cai$^{1,2}$,   Cong Li$^{1,2}$, Hao Chen$^{1,2}$, Lin Zhao$^{1}$,  Guodong Liu$^{1,2,4}$, Changqing Jin$^{1,2,4*}$, Zuyan Xu$^{5}$, Tao Xiang$^{1,2,3,4}$  and X. J. Zhou$^{1,2,3,4*}$}

\affiliation{
\\$^{1}$National Lab for Superconductivity, Beijing National laboratory for Condensed Matter Physics, Institute of Physics,
Chinese Academy of Sciences, Beijing 100190, China
\\$^{2}$University of Chinese Academy of Sciences, Beijing 100049, China
\\$^{3}$Beijing Academy of Quantum Information Sciences, Beijing 100193, China
\\$^{4}$Songshan Lake Materials Laboratory, Dongguan 523808, China
\\$^{5}$Technical Institute of Physics and Chemistry, Chinese Academy of Sciences, Beijing 100190, China
\\$^{\dag}$These authors contribute equally to the present work.
\\$^{\ddag}$Present address: Department of Materials Science and Engineering, Cornell University, Ithaca, New York 14853, USA
\\$^{*}$Corresponding author: XJZhou@iphy.ac.cn and Jin@iphy.ac.cn
}

\date{\today}
\maketitle
\newpage

{\bf High temperature superconductivity in cuprates arises from doping a parent Mott insulator by electrons or holes. A central issue is how the Mott gap evolves and the low-energy states emerge with doping. Here we report angle-resolved photoemission spectroscopy measurements on a cuprate parent compound by sequential {\it\textbf{in situ}} electron doping. The chemical potential jumps to the bottom of the upper Hubbard band upon a slight electron doping, making it possible to directly visualize the charge transfer band and the full Mott gap region. With increasing doping, the Mott gap rapidly collapses due to the spectral weight transfer from the charge transfer band to the gapped region and the induced low-energy states emerge in a wide energy range inside the Mott gap. These results provide key information on the electronic evolution in doping a Mott insulator and establish a basis for developing microscopic theories for cuprate superconductivity.
}\\


In high temperature cuprate superconductors,  the CuO$_2$ plane is a common and key structural element that  dictates their electronic structure and unusual physical properties.  In the undoped parent compound, each Cu site in the CuO$_2$ plane  has 9 electrons in the Cu 3{\it d} orbitals, which is expected to give a half-filled band and a metallic state according to the traditional band theory\cite{MattheissPRL,JFink1989}. Instead, it becomes an insulator because of the strong electron-electron correlation that splits the initial half-filled band into the lower Hubbard band (LHB) and  upper Hubbard band (UHB) separated by the on-site Coulomb repulsion U (Fig. 1a)\cite{JZaanen1985}.  The hybridization of the lower Cu 3{\it d} states and O 2{\it p} states gives rise to a charge transfer band (CTB) in between the lower and upper Hubbard bands, resulting in a correlated insulator with a charge transfer gap $\Delta$ (Fig. 1a)\cite{JZaanen1985}.  The cuprate parent compound is therefore, strictly speaking, a charge transfer insulator\cite{JZaanen1985} although it is generally believed that it can be described by an effective Hubbard Hamiltonian where the charge transfer band substitutes for the lower Hubbard band and the charge transfer gap $\Delta$  plays the role of U (Fig. 1a)\cite{PLeeRMP2006,ArmitageRMP2010}.

High temperature superconductivity in cuprates is realized by doping the CuO$_2$ planes with charge carriers (holes or electrons)\cite{PLeeRMP2006,ArmitageRMP2010}.  A key question is how the insulating state of the parent compound is transformed into a conductive or superconducting state upon carrier doping.  Such information on the electronic structure of the parent compound and its doping evolution is essential for establishing theories and understanding the origin of high temperature superconductivity\cite{EDagotto1994}.  Specifically, prominent issues are related to how the Mott gap evolves and how the low-lying electronic states emerge with doping\cite{HMatsumoto1989,Allen90,Meinders93,Veenendaal1994}.  The tunneling experiment, owing to its advantage in  measuring both the occupied state and unoccupied state,  has provided some key insight on the Mott gap in the parent compound\cite{CYeParent2013,WRuanParent} and  the emergence of low-lying states inside the gap upon slight carrier doping\cite{YKohsakaCCOCNP,PCaiBi2201NP16,YZhongInfinite19}. However, because the tunneling experiment measures the local density of states, it lacks the momentum information of electrons that is required in describing the complete electronic structure\cite{CYeParent2013,WRuanParent,YKohsakaCCOCNP,PCaiBi2201NP16,YZhongInfinite19}.   Angle-resolved photoemission spectroscopy (ARPES), with its unique momentum-resolving power,  has played a key role in studying the electronic structure of high temperature cuprate superconductors\cite{ADamascelliRMP}. However, because ARPES can only measure the occupied state, and the chemical potential of the parent compound and hole-doped compounds lies inside the Mott gap (Fig. 1a),  the ARPES measurements have not been possible to probe the entire gapped region in studying the parent compounds\cite{BOWellsPRL1995,FRonningScience1998,FRonningPRB2003,ORoschPRL2005,CHuCPL2018} and the evolution of the Mott gap and the low-lying states in lightly hole-doped cuprates\cite{TYoshidaPRL2003,KShenPRL2004,KShenScience2005,YYPengNC2013}.  Little information has been available on the momentum-resolved electronic structure evolution of the lightly doped cuprates that can cover the full energy spectrum of the gapped region\cite{PArmitageNCCO2002}.

Recently, {\it in situ} alkali metal deposition provides an effective way to fine tune surface electron doping in correlated systems like cuprates\cite{Hossain2008,Fournier2010,Zhang2016}, iridates\cite{Comin2012,Kim2014,Kim2015} and iron-based superconductors\cite{Miyata2015,Wen2016,Kyung2016}. Yet, to our knowledge, there has been no report of ARPES experiment with alkali metal deposition on cuprate parent compounds. It is challenging for ARPES experiments on these insulators, because the measurement temperature usually needs to be high, say $\sim$100K for Ca$_2$CuO$_2$Cl$_2$\cite{FRonningScience1998}, to avoid sample charging which hinders sufficient amount of the deposited alkali metal from staying stable on the sample surface.

In this paper, we report our ARPES  study on the electronic structure evolution in doping Ca$_3$Cu$_2$O$_4$Cl$_2$ (CCOC326), a double-layer parent compound of  high temperature cuprate superconductors. Fine tuning of the electron doping is realized by {\it in situ} depositing Rb onto the sample surface.  As soon as the CCOC326 parent compound is slightly electron doped, we find that the chemical potential  jumps from the position inside the Mott gap to the bottom edge of the upper Hubbard band. This makes it possible for ARPES to see the full Mott gap,  study its doping evolution and keep track on the doping-induced emergence of the low-lying electronic states.  The observed Mott gap represents an indirect gap of $\sim$1.5 eV between the charge transfer band top at $(\pi/2,\pi/2)$ and the upper Hubbard band bottom at $(\pi,0)$.  With electron doping, the Mott gap shows little change in its magnitude but it rapidly collapses due to the spectral weight depletion of the charge transfer band and the emergence of low-lying states in the gapped region.  Distinct doping evolution of the low-energy states is observed between the $(\pi/2,\pi/2)$ nodal region and the $(\pi,0)$ Brillouin zone face.  Upon a slight electron doping, the upper Hubbard band is filled with electrons and a metallic electron-like band develops around the $(\pi,0)$ zone face. In the meantime, near the  $(\pi/2,\pi/2)$ nodal region, broad in-gap states emerge inside the gapped region spreading over a wide energy range that approach the chemical potential with increasing doping. These results provide momentum-resolved information on the electronic structure evolution in lightly doping a Mott insulator over the entire gapped region and establish a basis for developing microscopic theories in understanding unusual physical properties and high temperature superconductivity in cuprate superconductors.

\vspace{5mm}
\noindent {\bf\large Results}\\
The Ca$_3$Cu$_2$O$_4$Cl$_2$ represents a prototypical parent compound of high temperature cuprate superconductors\cite{TSowa1990}. Its crystal structure consists of two CuO$_2$ planes in one structural unit (Fig. 1b)  and  is analogous to Ca$_2$CuO$_2$Cl$_2$ (CCOC214) with one CuO$_2$ plane in a unit cell\cite{FRonningScience1998,CHuCPL2018}.  We choose to work on this CCOC326 parent compound because it is less insulating than its single-layer counterpart CCOC214\cite{JFZhao2019}, so that our ARPES measurements can be performed at lower temperature without sample charging problem (see Supplementary Fig. 1).  The electronic structure of  CCOC326 (Supplementary Fig. 2) is similar to that of CCOC214\cite{FRonningScience1998,CHuCPL2018} but with a smaller energy gap and sharper spectral features. We deposit Rb onto the sample surface to fine-tune the electron doping level (Fig. 1c); this can realize systematic ARPES measurements on the same sample under the same experimental condition.  The lower measurement temperature of CCOC326 is crucial for the success of our present work because only at low temperature can sufficient amount of Rb be absorbed on the sample surface and stay stable to provide high enough electron doping.

Figure 1d shows the Fermi surface mapping of Ca$_3$Cu$_2$O$_4$Cl$_2$ at different Rb deposition stages.  Little spectral weight is observed for the initial two stages (1 and 2), which is consistent with the insulating nature of the CCOC326 parent compound (see Supplementary Fig. 2).  The spectral weight starts to appear after stage 3 near the (0,$\pi$) and (-$\pi$,0) regions in the covered momentum space. It gets stronger with further Rb deposition till stage 8. Then an electron-like Fermi pocket is observed for stage 9 and its size increases with Rb deposition. For the stage 11, the spectral weight near the nodal region starts to emerge and gets stronger with further Rb deposition.  The overall Fermi surface evolution with Rb deposition is similar to the previous ARPES measurements on typical electron-doped cupates (Nd,Ce)$_2$CuO$_4$\cite{PArmitageNCCO2002},  confirming that electron doping is realized in the CCOC326 parent compound by Rb deposition.   By the Luttinger sum rule\cite{PArmitageNCCO2002,DJSongPRL2016},  the electron doping level p is calculated from the observed Fermi surface area for the later deposition stages, and estimated for the initial stages (1 and 2) from the linear extrapolation according to the deposition procedures, as shown in Fig. 1e.  We see that the doping level is well controlled under the sequential Rb deposition, in particular, it covers the lightly doped region (p$\sim$0-0.04) which is the key range to keep track on the electronic structure evolution from the parent compound.  This regime is the main focus of the present work that was not covered in the previous ARPES measurements\cite{PArmitageNCCO2002,DJSongPRL2016,ArmitageRMP2010}.  Considering that CCOC326 can also be doped with holes to reach superconductivity\cite{CQJinNature1995,YZenitani1995},  our present work indicates that this double-layer cuprate parent compound Ca$_3$Cu$_2$O$_4$Cl$_2$ is a new ambipolar cuprate system that can be doped by either holes or electrons, in addition to limited cases found in other cuprate superconductors\cite{KSegawa2006,YZhongInfinite19}.

Figure 2 shows the band structure evolution with Rb deposition measured along three high symmetry momentum cuts,  (-$\pi$,-$\pi$)-($\pi$,$\pi$) nodal direction (Fig. 2a),  (-$\pi$,0)-(0,$\pi$) cut crossing both the (-$\pi$/2,-$\pi$/2) nodal point and (0,$\pi$) antinodal point (Fig. 2b), and (-$\pi$,-$\pi$)-(-$\pi$,$\pi$) cut along the zone face (Fig. 2c). For the undoped CCOC326, its band structure is characterized by the charge transfer band that shows a clear dispersion along the nodal direction with its top $\sim$300 meV below the Fermi level  (leftmost panel in Fig. 2a) and a nearly vertical band along the zone face with its top $\sim$600 meV below the Fermi level (leftmost panel in Fig. 2c) (see also Supplementary Fig. 2).  As soon as the sample is doped with a tiny amount of electrons (less than 0.005 for stages 1 and 2 in Fig. 2), both bands (second and third panels from left  in Fig. 2a and Fig. 2c) abruptly sink down by $\sim$1.1 eV.  As will be discussed later, this is due to the chemical potential jump to the bottom edge of the upper Hubbard band.  This makes it possible for ARPES to directly access the entire Mott gap in the cuprate parent compound, which is an indirect gap defined by the top of the charge transfer band at the ($\pi$/2,$\pi$/2) nodal point and the bottom of the upper Hubbard band at the ($\pi$,0) antinodal point. The gap size thus determined is $\sim$1.5 eV. This agrees with the gap value measured directly by the tunneling experiments\cite{WRuanParent}. These results indicate that, right after a slight electron doping, the chemical potential shifts to the bottom edge of the upper Hubbard band. With further electron doping, more states are created in the upper Hubbard band, gradually forming a parabolic electron band near ($\pi$,0). In this case, the chemical potential is defined by this newly formed band in the upper Hubbard band. This is consistent with the usual picture that is proposed in literature\cite{ArmitageRMP2010,PArmitageNCCO2002}.

From Fig. 2,  we can directly visualize how the low-lying electronic states emerge and evolve inside the Mott gap with increasing electron doping.  Overall, with the introduction of electrons, new states appear inside the gap; their spectral weight get stronger with increasing electron doping.  Also the newly developed states spread over a wide energy range inside the gap region, indicating the incoherent nature of the correlated electronic states. In the meantime, the spectral weight of the charge transfer band decreases with the doping increase. In particular, the low-lying in-gap states exhibit distinct momentum-dependence; their behavior along the nodal direction (Fig. 2a) is markedly different from that along the zone face (Fig. 2c).  Along the zone face (Fig. 2c), the low-energy states appear starting from the Fermi level, then spread downward to high-binding energy.  With increasing doping, their spectral weight gets stronger. At high doping (stages 9$\sim$13 in Fig. 2c),  coherent parabolic electron-like bands form near the ($\pi$,0) region.  Note that, along the zone face, the newly developed in-gap states lie at the Fermi level as soon as they appear upon tiny electron doping. On the other hand, along the nodal direction (Fig. 2a), the in-gap states appear from the high-binding energy near the top of the charge transfer band, spreading upwards with the increase of the spectral weight with increasing electron doping.  However, the bands do not cross the Fermi level at lower doping but gradually approach the Fermi level with increasing doping.  The band reaches the Fermi level only at high doping (stages 12 and 13 in Fig. 2a). We find that the newly developed in-gap states along the nodal direction (Fig. 2a) have quite similar ``waterfall" band dispersion as the original charge transfer band. This is different for the in-gap states along the zone face where they exhibit quite different band dispersion as the original vertical bands. It is also interesting to notice that, the band evolution with electron doping along the nodal direction in the present work is reminiscent to that observed in lightly hole-doped cuprates\cite{TYoshidaPRL2003,KShenPRL2004,KShenScience2005,YYPengNC2013} although the behaviour near the ($\pi$,0) antinodal region is completely different between the lightly electron-doped and lightly hole-doped cases. In addition to the evolution of low-lying electronic states mentioned above, we also observe a profound spectral change in the high-binding energy region from 2 to 2.5 eV with increasing electron doping. The observation of such a strong modification is important because generally it is not expected to see such a strong change in such a high energy region with electron doping.

We also note that, when the deposited alkali metal is thick enough to form full layers, they may produce free-electron-like bands that are centered around the Brillouin zone center ($\Gamma$ point)\cite{Kim2014}. In our measurements, we do not observe any free-electron-like bands near the Brillouin zone center. Moreover, the observed in-gap states lie near the ($\pi$/2,$\pi$/2) nodal region and ($\pi$,0) antinodal region. This can fully rule out the possibility that the in-gap states originate from the Rb valence band; they represent the intrinsic features of the doped cuprate parent compounds.

Figure 3 shows how the charge transfer band and the low-lying in-gap states evolve with electron doping in terms of their positions and spectral weight.  Fig. 3a and 3b show the photoemission spectra (energy distribution curves, EDCs) at ($\pi$/2,$\pi$/2) nodal point and ($\pi$,0) antinodal point, respectively, at different Rb deposition stages. The EDCs at ($\pi$/2,$\pi$/2) nodal point (Fig. 3a) are characterized by two main features: one is the well-defined peak (marked by solid circles in Fig. 3a)  from the charge transfer band which loses its weight with increasing doping, and the other is the broad feature inside the gapped region between [-1,0] eV forming the low-lying electronic states that gain weight with increasing doping.  For the  EDCs at ($\pi$,0) antinodal point (Fig. 3b), in addition to the feature  near -1.5 eV in the doped samples that is related to the charge transfer band, the main feature is the in-gap states that grow up in intensity with increasing doping and form a peak marked by triangles in Fig. 3b. Fig. 3c  shows the energy position of the EDC peaks at ($\pi$/2,$\pi$/2)  (marked by solid circles in Fig. 3a) related to the top position of the charge transfer band, and that of the EDC peaks at ($\pi$,0) ( marked by triangles in Fig. 3b) related to the bottom position of the upper Hubbard band. Both positions show a jump right after the electron doping, and then exhibit a gradual shift to higher binding energy with increasing electron doping (Fig. 3c).  Their energy difference (Fig. 3d) represents the Mott gap magnitude that shows little change with the doping level and stays around $\sim$1.5 eV. On the other hand, the EDCs at ($\pi$/2,$\pi$/2), which is the top of the charge transfer band, lose spectral weight rapidly with initial electron doping (Fig. 3a and 3e); the spectral weight of the charge transfer band nearly disappears at a doping level of $\sim$0.04 (Fig. 3e).  The spectral weight of the in-gap states, either at ($\pi$/2,$\pi$/2) or  ($\pi$,0) points, keeps increasing with electron doping, as shown in Fig. 3(a,b) and Fig. 3e. These results indicate that, with electron doping, although the magnitude of the Mott gap shows little change, the gap collapses due to the vanishing of the initial charge transfer band, and the filling-up of the in-gap states across the entire gapped region.

One key issue in doping the parent compounds of the cuprate superconductors is how the chemical potential shifts with doping; whether it is pinned inside the gapped region to form the confined in-gap states or it shifts to the top of the charge transfer band or the bottom of the upper Hubbard band upon respective hole- or electron-doping\cite{HMatsumoto1989,Allen90,Meinders93,Veenendaal1994,ADamascelliRMP,PLeeRMP2006,ArmitageRMP2010}. Our detailed doping evolution of the band structure (Fig. 2) clearly indicates that, upon tiny electron doping, the chemical potential $\mu$, which lies close to the charge transfer band in the parent compound, shifts immediately to the bottom of the upper Hubbard band.
Photoemission can measure the relative shift of the chemical potential as a function of carrier doping with reference to the electronic states located far away from the Fermi level thus their energies should be qualitatively unaffected by doping\cite{KShenPRL2004,MIkeda2010}.
To this end, we carried out photoemission measurements on high-binding energy O 2$p$ nonbonding states which show up as sharp peaks at $\Gamma$ (Fig. 4a) and at ($\pi$,$\pi$) point (Fig. 4b), and the Ca 3$p$ core level near the binding energy of $\sim$25 eV (Fig. 4c) in CCOC326 at different Rb deposition stages.  A consistent result on the relative chemical potential shift with electron doping in CCOC326 is obtained in Fig. 4d by keeping track on the relative EDC peak position change with Rb deposition from the three cases in Fig. 4(a,b,c). A chemical potential jump of $\sim$1.0 eV is observed right after a tiny amount of electrons are introduced into the parent compound (Fig. 4d). We note that since the parent compound is a perfect insulator, the chemical potential is not well-defined and may vary its position inside the gap between different samples.  The magnitude of the chemical potential jump may vary slightly among different samples, but the jump itself is intrinsic because it will shift to the bottom of the upper Hubbard band as soon as the electrons are doped. As seen in Fig. 4d, upon further electron doping in CCOC326, the chemical potential gradually increases, consistent with the chemical potential shift in other electron-doped cuprates\cite{MIkeda2010}.

It is important to investigate, after Rb deposition and some electrons are transferred into the CuO$_2$ planes, whether there is a band bending that may affect the chemical potential determination. We think the situation in Rb-deposited CCOC326 is different from the usual band-bending picture. Before Rb deposition, the CuO$_2$ layer, Ca layer and CaCl layer already have charge of 2-, 2+ and 1+, respectively, that form internal electrical field in CCOC326. The charge transfer of 2$\delta$ electrons from Rb deposition layer to the two CuO$_2$ planes only slightly modifies the charge distribution. As seen in our experiments, dramatic changes of the chemical potential shift and band structure occur before $\delta$$\sim$0.04.  The corresponding charge change of CuO$_2$ plane from 2- to (2+0.04)- is small, only accounting for 2$\%$ change. It is expected that in this case, the chemical potential shift is mainly caused by the electron doping of the CuO$_2$ plane, but not by the band bending near the surface. In fact, there is no band bending between the Rb deposition layer and the adjacent CaCl layer because they both have positive charge. This is consistent with the fact the chemical potential shift obtained from Ca 3$p$ in CaCl layer is consistent with that obtained from O 2$p$ in CuO$_2$ plane although they have different internal electric field environment. The chemical potential shift we observed is also consistent with that observed in other electron-doped cuprates in bulk form\cite{MIkeda2010}.

In ARPES study of YBa$_2$Cu$_3$O$_{6.5}$, \emph{in situ} potassium deposition has been successfully applied to continuously tune the surface doping from heavily overdoped to underdoped regime\cite{Hossain2008} and further into heavily underdoped regime\cite{Fournier2010}. CCOC326 is similar to YBa$_2$Cu$_3$O$_{6.5}$ in that they both have two CuO$_2$ planes in one structural unit. But the charge transfer distance in CCOC326 is shorter than that in YBa$_2$Cu$_3$O$_{6.5}$ because the former crosses one CaCl layer while the latter needs to cross two CuO$_x$ and BaO layers. Since ARPES is a surface sensitive technique, at 21.2 eV photon energy, the probe depth is on the order of $\sim$10 {\AA} that matches the surface doped region so the doping-induced change of electronic structure near the surface can be fully detected by ARPES. The ARPES measurements of K-deposited YBa$_2$Cu$_3$O$_{6.5}$ system\cite{Hossain2008,Fournier2010} and Bi$_2$Sr$_2$CaCu$_2$O$_{8+\delta}$\cite{Zhang2016} have provided a direct evidence on measuring genuine doping evolution of bulk cuprates. The consistency of our results at higher doping with the ARPES measurements on bulk samples\cite{PArmitageNCCO2002} also indicates that our results represent genuine doping evolution of bulk cuprates.

The doping-induced band structure evolution (Figs. 2 and 3), combined with the shift of the chemical potential (Fig. 4), provides a momentum-resolved electronic structure picture in electron doping the cuprate parent compound. Fig. 5a and 5b show the band structure measured along (-$\pi$,-$\pi$)-($\pi$,$\pi$) nodal direction and (-$\pi$,-$\pi$)-(-$\pi$,$\pi$) zone face, respectively, at several representative doping levels.
Here the relative energy scale is used and all data are plotted with reference to the chemical potential $\mu_0$ using the values from Fig. 4d, thus highlighting the change of the chemical potential with doping.
In this case, it is also straightforward to visualize the electronic structure variation with doping in three energy regions: charge transfer band (CTB) region, in-gap states (IGS) inside the energy gap region, and the upper Hubbard band (UHB) region, as marked on the top of Fig. 5a.   Fig. 5c and 5d show the momentum-integrated EDCs obtained from Fig. 5a and 5b, respectively. By summing up all the spectral weight over the covered momentum space, the integrated EDC represents the local density of states along the particular momentum cut. With electron doping, the spectral weight transfer from the charge transfer band to the in-gap states is clear, with the spectral weight depletion of the charge transfer band and the increase of the in-gap states.

Combining the results in Fig. 5(a-d), one gets a schematic electronic structure evolution picture in a momentum-resolved manner, as shown in Fig. 5e and Fig. 5f for the (-$\pi$,-$\pi$)-($\pi$,$\pi$) nodal cut and (-$\pi$,-$\pi$)-(-$\pi$,$\pi$) cut, respectively.  A few key points emerge in electron doping of the parent Mott insulator.  First,  the chemical potential jumps from the in-gap position to the bottom of the upper Hubbard band right after a tiny amount of electron doping;  Second, emergence of low-lying electronic states which span a wide energy range inside the gapped region; Third, the Mott gap magnitude changes little with doping but the gap collapses due the depletion of the charge transfer band and the filling of the low-lying electronic states in the gapped region. Here we note that, due to the Fermi cutoff, we have no access to the entire upper Hubbard band although one would expect that the upper Hubbard band also loses its weight with increasing electron doping\cite{HKusunose2003,CYeParent2013}. Fourth, momentum anisotropy of the low-energy in-gap states. A metallic state forms near the  ($\pi$,0)  antinodal region for the lightly doped sample, but the state near the ($\pi$/2,$\pi$/2) nodal region is away from the Fermi level when it is formed. It gradually approaches the Fermi level with increasing doping and only crosses the Fermi level at higher doping level above p$\sim$0.1.

\vspace{5mm}
\noindent {\bf\large Discussion}\\
The electronic structure of the parent compound and the lightly doped cuprates is of fundamental importance to understanding the high temperature superconductivity mechanism and has attracted enormous theoretical attention\cite{EDagotto1994,AGeorges1996,MImada1998,PLeeRMP2006,ArmitageRMP2010,PPhillips2010}.  The basic theoretical model for describing the half-filled cuprates,  with a single electron on the Cu $d_{x^2-y^2}$ orbital and filled O 2$p_x$ and 2$p_y$ orbitals,  is the three-band Hubbard model that considers electron hopping (t),  the on-site Coulomb interaction (U$_d$), the energy difference between oxygen and copper orbitals ($\Delta$$_{pd}$), and the intersite interaction (V$_{pd}$)\cite{VJEmery1987,CMVarma1987}.  Anderson proposed that a single-band Hubbard model containing the hopping integral between the nearest-neighbour sites (t) and the on-site Coulomb repulsion (U) can describe the cuprate parent compound\cite{PAnderson1987}.  In the case where U$\gg$t as in the cuprates, the single-band Hubbard model can be further simplified into one-band t-J model which is a possible minimal model for the cuprates\cite{FCZhang1988}.  Since the hopping process may also involve the second (t$^{'}$)  and third (t$^{''}$) nearest neighbours, an extended t-J model, the t-t$^{'}$-t$^{''}$-J model, was also proposed\cite{TTohyama2000}. It remains controversial whether the three-band Hubbard model can be simplified into single-band Hubbard model\cite{VJEmery1988,HEskes1991,CMVarma1997}. Also even the single-band Hubbard model or t-J model cannot be solved exactly and approximations have to be applied in the computations.
Theoretically, the single-particle spectrum has been mainly calculated from the single-band Hubbard or t-J model. These calculations provide a big picture that agrees well with our measurements on some aspects like the Mott gap structure in the parent compound and emergence of in-gap states with doping. In this sense, we think that the overall electronic structure evolution can be described by the single-band Hubbard model. However, no solutions of the Hubbard model can reproduce the coexistence of fairly incoherent doped states with large band width extending almost across the charge transfer gap with a very dispersive lower Hubbard band, as observed in our present experiments.
Therefore, our momentum-resolved electronic structures on the parent compound and lightly doped cuprates covering the entire Mott gap region provide key information to examine on these microscopic theoretical models, particularly on those that are applied to electron-doped cuprates\cite{CKusko2002,HKusunose2003,TTohyama2004,TXiangPRB2009,CWeber2010}.

Our ARPES results reported here touch several central issues on the electronic structure evolution of doping Mott insulators in cuprates. The first is about the nature of the Mott gap. It was suggested that the Mott gap is an indirect gap in cuprate parent compounds\cite{TTohyama2001,CKusko2002}.  Optical\cite{SUchida1991,TArima1993,KWaku2004} and tunneling\cite{CYeParent2013,WRuanParent} measurements detected the Mott gap in the parent compounds but cannot distinct whether it is a direct or indirect gap.  By spanning the full gapped region, our momentum-resolved results on the nearly undoped CCOC326 provide direct evidence that the Mott gap is indeed an indirect gap from the top of the charge transfer band at ($\pi$/2,$\pi$/2) to the bottom of the upper Hubbard band, consistent with the results from (Nd,Ce)$_2$CuO$_4$\cite{PArmitageNCCO2002}. The second issue concerns the chemical potential shift upon doping. It has been under debate whether in the lightly electron-doped cuprates, the chemical potential will be pinned inside the gapped region or it will shift to the bottom of the upper Hubbard band\cite{Meinders93,AGeorges1996}.   Previous results on electron-doped (Nd,Ce)$_2$CuO$_4$ did not observe the chemical potential jump in lightly doping the parent compound\cite{NHarima2001,PArmitageNCCO2002}.  This may be related to the possibility that the parent Nd$_2$CuO$_4$ sample is not truly undoped. With CCOC326 being a perfect insulator at zero doping, our present work provides direct evidence that, in the parent cuprate compound, upon a tiny electron doping, the chemical potential will jump to the bottom edge of the upper Hubbard band.  The third issue is about the nature of the newly-formed in-gap states. It has been discussed whether or not the in-gap states are coherent quasiparticle states that are confined in a narrow energy range\cite{CKusko2002,HKusunose2003}. Our present work indicates that the newly-developed in-gap states are incoherent that spread over a large energy range inside the gapped region.  Along the nodal direction, although the in-gap states form a well-defined waterfall-like band (Fig. 2a),  the photoemission spectra (EDCs) are broad that spread over a wide energy range. Along the zone face near the antinodal region, although only a narrow band is formed around ($\pi$,0) with elelctron doping (Fig. 2c), the photoemission spectra are broad (Fig. 3b) and the spectral tails extend to a wide energy range.  This incoherence of the in-gap states may be related to the strong correlation effect in the lightly doped cuprates.  The fourth issue is about the Mott gap. It has been under debate whether, with carrier doping, the Mott gap will change its size or not\cite{CKusko2002}. Our present work indicates that, the Mott gap magnitude changes little with electron doping (Fig. 3d). But it collapses because of the spectral weight depletion of the charge transfer band and the filling-up of low-lying states inside the gap (Fig. 3(a,b,e)).

Our momentum-resolved results on lightly electron-doped  CCOC326 over the entire gapped region show that the overall electronic structure evolution can be described by the single-band Hubbard model, as shown in Fig. 5(e,f).  However, the in-gap states behave not as coherent quasiparticles, but as incoherent states that extend over a wide energy distribution. In particular, the in-gap states are anisotropic in momentum space; the electronic evolution near the ($\pi$/2,$\pi$/2) nodal region is distinct from that near the ($\pi$,0) antinodal region. The appearance of waterfall-like in-gap states with a large bandwidth along the nodal direction, its position lying below the Fermi level at lower doping, and the gradual approaching of the band towards the Fermi level with increasing doping,  cannot be explained by the existing theories\cite{CKusko2002,HKusunose2003,TTohyama2004,TXiangPRB2009,CWeber2010}. Our present results ask for further theoretical efforts to propose new theories to understand the electronic structure evolution in lightly doped cuprates, in addition to discriminate on the existing theoretical models.

\vspace{3mm}
\noindent {\bf\large Methods}\\
\noindent {\bf Sample preparation}\\
High quality Ca$_3$Cu$_2$O$_4$Cl$_2$ single crystals were grown by a self-flux method with two steps\cite{JFZhao2019}.
Firstly, polycrystalline Ca$_3$Cu$_2$O$_4$Cl$_2$ samples were prepared by conventional solid-state reaction method using high-purity raw materials CaO (Alfa, 99.95\% pure), CuCl$_2$ (Alfa, 99.995\% pure) and CuO (Alfa, 99.995\% pure). The powder mixture of CaO, CuCl$_2$ and CuO with a molar ratio of 3:1:1 was ground together and heated at 800 $^\circ$C for 20 hours with several intermediate grindings in order to obtain single-phase samples. Secondly, polycrystalline Ca$_3$Cu$_2$O$_4$Cl$_2$ samples were put into an alumina crucible and heated to 1060 $^\circ$C with the rate of 180 $^\circ$C/hour, dwelled at 1060 $^\circ$C for 10 hours and cooled to 900 $^\circ$C with the rate of 30 $^\circ$C/hour. The furnace was shut off at 900 $^\circ$C to let the samples cool rapidly in order to avoid generating Ca$_2$CuO$_2$Cl$_2$ phase.

\noindent {\bf High-resolution ARPES measurements}\\
High-resolution angle-resolved photoemission measurements were carried out on our lab-based system\cite{GDLiu2008,XJZhou2018}  that is equipped with a Scienta Omicron DA30L electron analyzer. A Scienta Omicron helium discharge lamp with photon energies of 21.2 eV and 40.8 eV was used as the light source. The energy resolution was set at 10$\sim$20 meV for Fermi surface mapping and band structure measurements. The angular resolution was $\sim$0.3$^\circ$.  Samples were cleaved at 80 K and measured at 55 K to avoid sample charging (see Supplementary Fig. 1).   A SAES dispenser is used to evaporate rubidium (Rb) {\it in situ} onto the Ca$_3$Cu$_2$O$_4$Cl$_2$ surface step by step.  ARPES measurements were carried out after each Rb deposition and such a process is repeated many times.  All measurements were carried out in ultrahigh vacuum with a base pressure better than 5$\times$10$^{-11}$ torr.

\vspace{3mm}
\noindent {\bf\large Data Availability}\\
The data that support the findings of this study are available from the corresponding author upon reasonable request.

\vspace{3mm}
\noindent {\bf\large Acknowledgement}\\
We thank financial support from the National Natural Science Foundation of China (Grant No. 11888101), the National Key Research and Development Program of China (Grant No. 2016YFA0300300 and 2017YFA0302900), the Strategic Priority Research Program (B) of the Chinese Academy of Sciences (Grant No. XDB25000000),  the Youth Innovation Promotion Association of CAS (Grant No. 2017013), and the Research Program of Beijing Academy of Quantum Information Sciences (Grant No. Y18G06).

\vspace{3mm}
\noindent {\bf\large Author contributions}\\
X.J.Z. and C.H. proposed and designed the research. C.H. carried out the ARPES experiments with Q.G. and H.T.Y.. J.F.Z. and C.Q.J. grew the Ca$_3$Cu$_2$O$_4$Cl$_2$ single crystals. C.H., Q.G., H.T.Y., H.T.R., J.W.H., J.L., Y.Q.C., C.L., H.C., L.Z., G.D.L., Z.Y.X. and X.J.Z. contributed to the development and maintenance of Laser-ARPES system. T.X. contributed to theoretical analysis.  C.H. and X.J.Z. wrote the paper. All authors participated in discussions and comments on the paper.

\vspace{3mm}
\noindent {\bf\large Competing interests}\\
The authors declare no competing interests.

\vspace{3mm}
\noindent {\bf\large Additional information}\\

\newpage

\begin{figure*}[tbp]
\begin{center}
\includegraphics[width=1.0\columnwidth,angle=0]{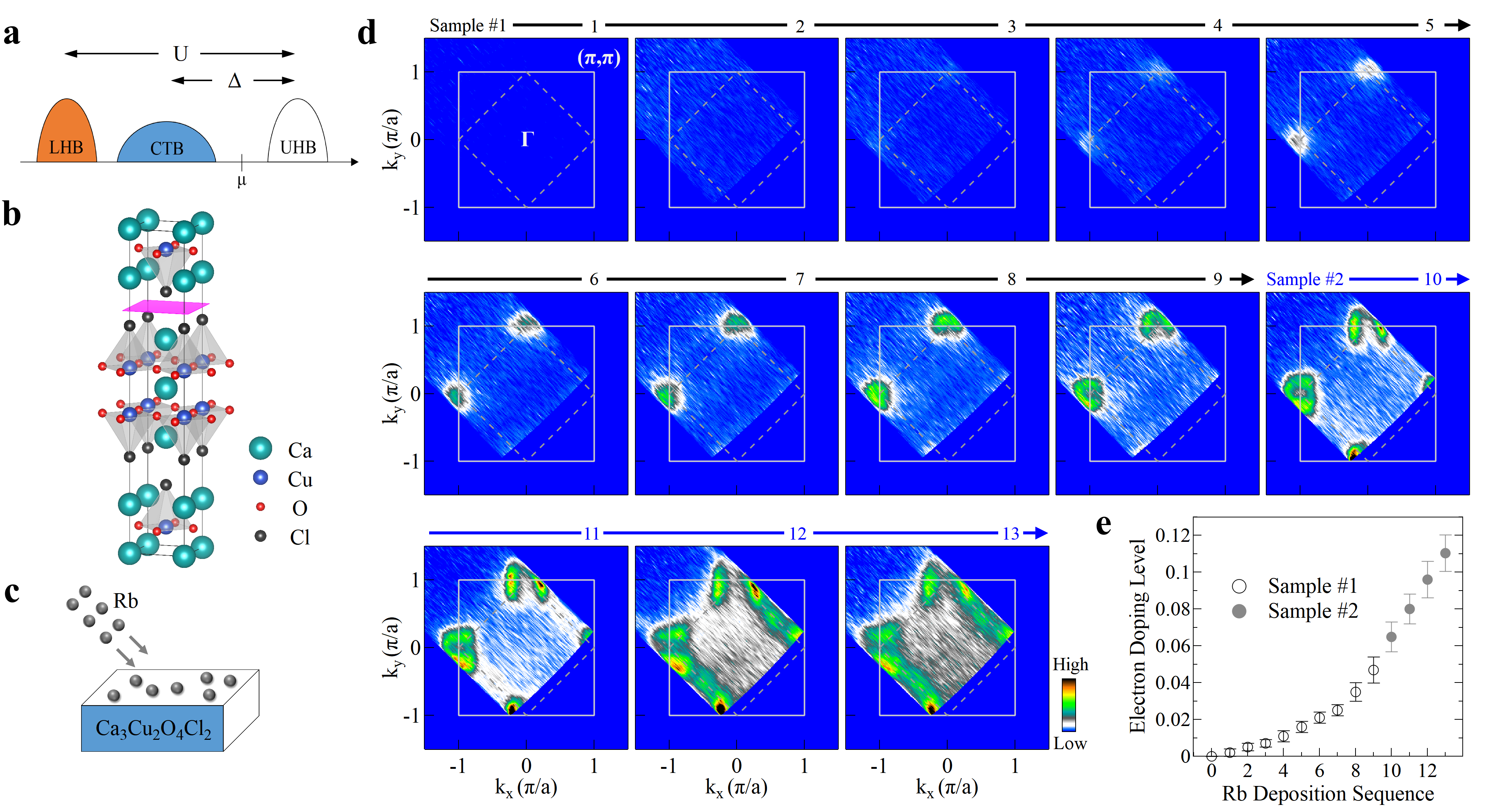}
\end{center}
\caption{{\bf Fermi surface evolution of Ca$_3$Cu$_2$O$_4$Cl$_2$ with Rb deposition.} (a) Schematic electronic structure picture of the cuprate parent compound. It consists of lower Hubbard band (LHB) and upper Hubbard band (UHB) that are separated by the on-site Coulomb repulsion U.  The charge transfer band (CTB) lies in between the LHB and UHB bands, giving rise to a charge transfer gap $\Delta$. The chemical potential $\mu$ is located inside the gapped region.  (b) Crystal structure of CCOC326 consisting of  double CuO$_2$ planes in one structural unit. The crystal cleaves between two adjacent CaCl layers, as marked by the pink sheet. (c) Schematic illustration of \emph{in situ} Rb deposition onto the sample surface. Electrons are transferred into the top layers due to the physical absorption of Rb on the sample surface. (d) Fermi surface mapping of CCOC326 at different Rb deposition stages. For convenience, we use 0$\sim$13 to denote the deposition sequence where 0 represents the original sample before Rb deposition. The images for sequences 0$\sim$9 are obtained on one sample while that for sequences 10$\sim$13 are obtained from another sample which complement each other. The Fermi surface images are obtained by integrating the spectral weight over an energy window of [-0.03, 0.03] eV with respect to the Fermi level.  (e) Estimated electron doping level  after each Rb deposition stage. For the deposition stages from 3 to 11, the electron doping level is calculated from the area of the electron-like Fermi surface centered at (-$\pi$,0). For the deposition sequences 12 and 13, it is estimated from the area of the hole-like Fermi surface centered at (-$\pi$,$\pi$). The doping level for the deposition  stages 1 and 2 are obtained by linear extrapolation in the low doping region according to the deposition procedures. Error bars reflect the uncertainty in determining the Fermi momentum.
}
\end{figure*}

\begin{figure*}[tbp]
\begin{center}
\includegraphics[width=1.0\columnwidth,angle=0]{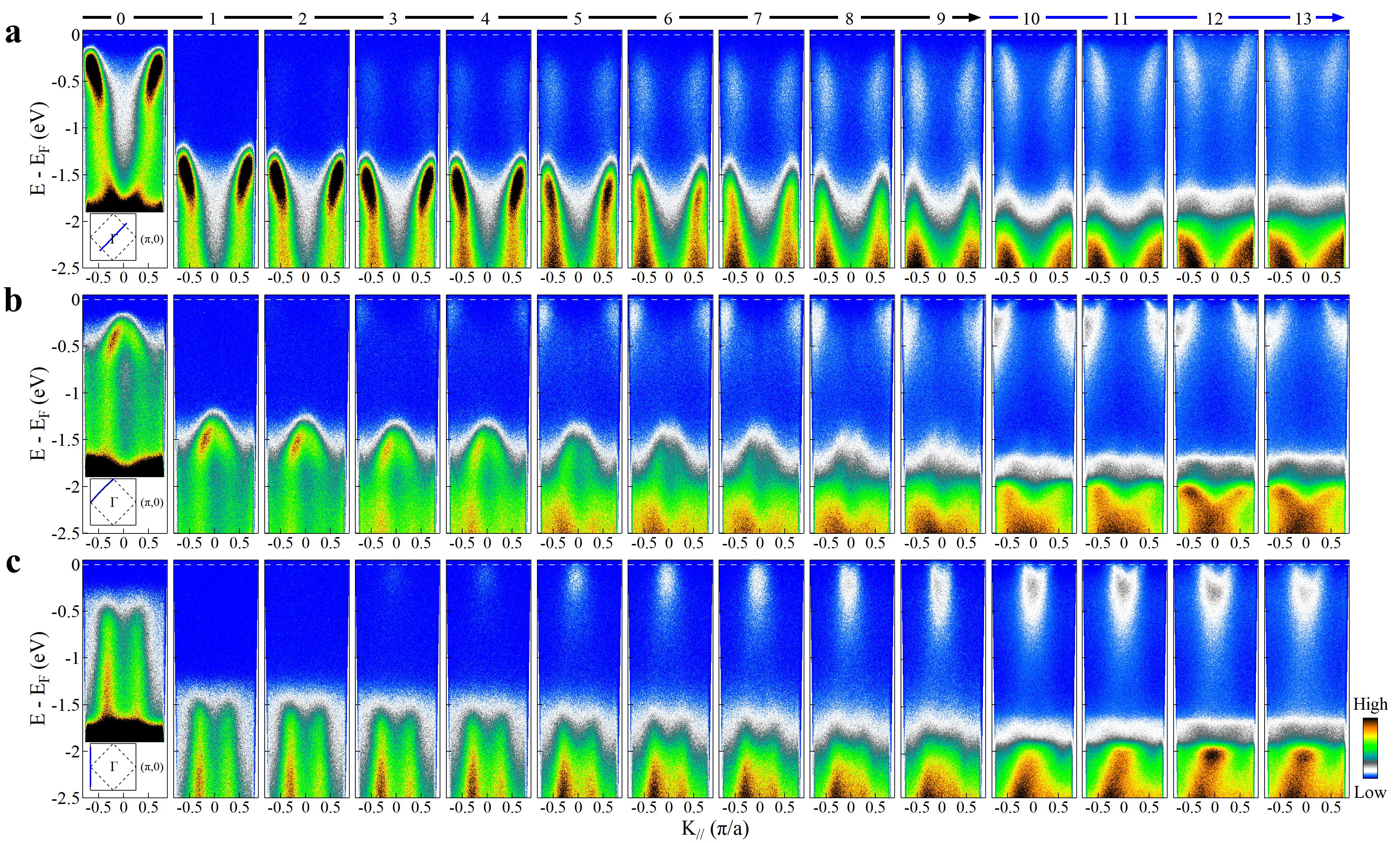}
\end{center}
\caption{{\bf Band structure evolution of Ca$_3$Cu$_2$O$_4$Cl$_2$ with Rb deposition.} (a) Doping evolution of band structure along the (-$\pi$/2,-$\pi$/2)-($\pi$/2,$\pi$/2) nodal momentum cut.  (b) Band structure evolution with Rb deposition along the (-$\pi$,0)-(0,$\pi$) momentum cut.  (c) Band structure evolution with Rb deposition along the  (-$\pi$,-$\pi$)-(-$\pi$,$\pi$)  cut. The locations of the three momentum cuts are illustrated by the blue solid line in the inset of panels (a), (b) and  (c), respectively.
}
\end{figure*}

\begin{figure*}[tbp]
\includegraphics[width=1.0\columnwidth,angle=0]{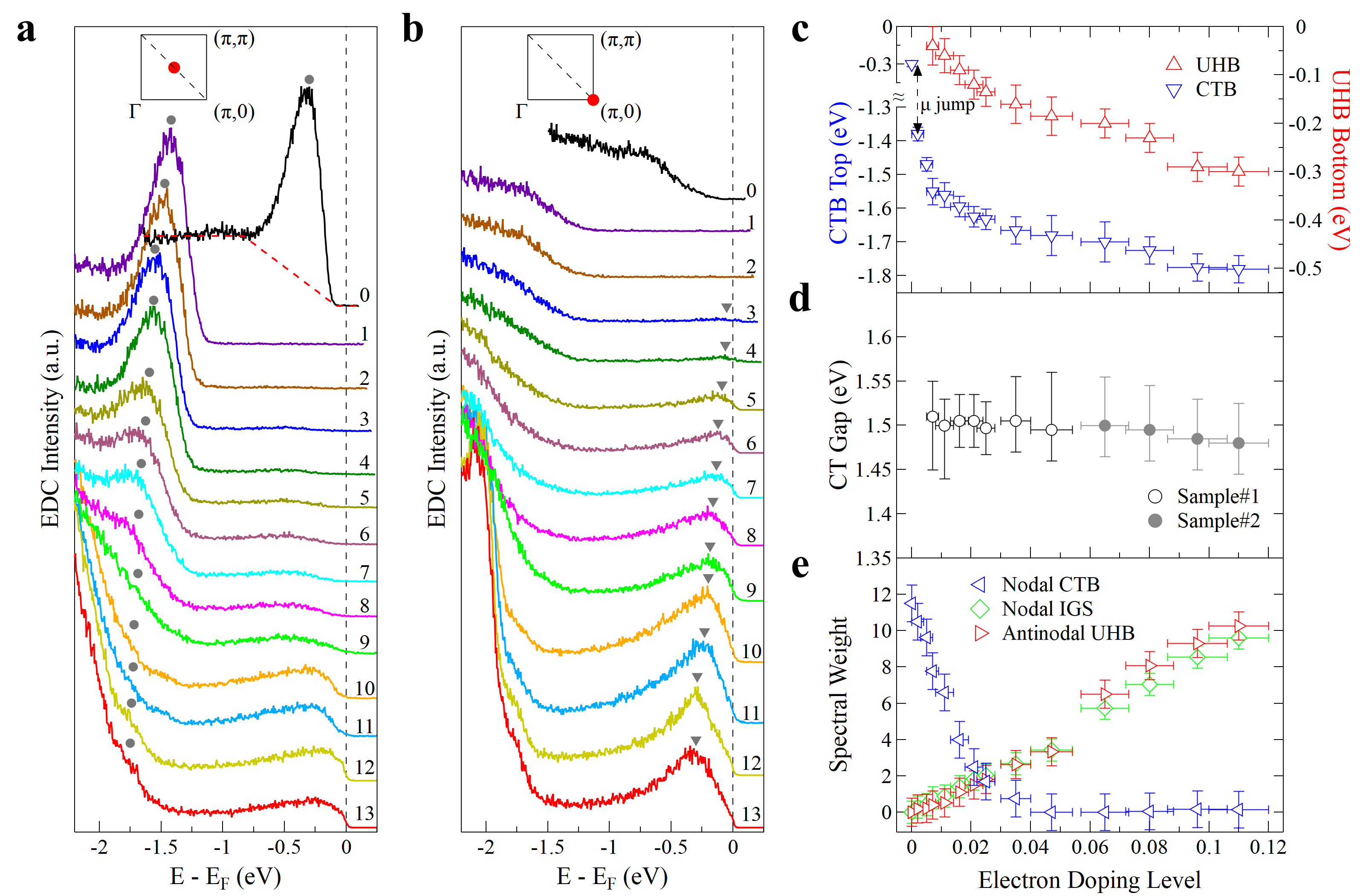}
\begin{center}
\caption{{\bf Doping evolution of the charge transfer band (CTB), upper Hubbard band (UHB) and the low-energy in-gap states (IGS).}  (a) Doping evolution of EDCs taken at ($\pi$/2,$\pi$/2) point. The peak position of the EDCs, marked by grey circles, defines the top of the nodal charge transfer band.  (b) Doping evolution of EDCs taken at ($\pi$,0) point.  The peak position of the EDCs, marked by grey triangles, represents the bottom of the antinodal upper Hubbard band.  (c)  The EDC peak positions of the nodal charge transfer band and antinodal upper Hubbard band extracted from (a) and (b), respectively.   (d)  The magnitude of the charge transfer gap (CT Gap) determined from the difference of the nodal CTB and antinodal UHB positions in (c).  (e) Doping evolution of the spectral weight of the nodal CTB (blue triangles), antinodal UHB (red triangles), and nodal in-gap states (IGS) (green diamonds). The nodal CTB spectral weight is obtained from the EDC peak area in (a) after subtracting the linear background as illustrated by the red dashed line in (a) for the EDC of the undoped sample. The spectral weight of antinodal UHB and nodal IGS are obtained by integrating the EDCs in (b) and (a) over an energy window [-1, 0.03]eV, respectively.
}
\end{center}
\end{figure*}

\begin{figure*}[tbp]
\begin{center}
\includegraphics[width=1\columnwidth,angle=0]{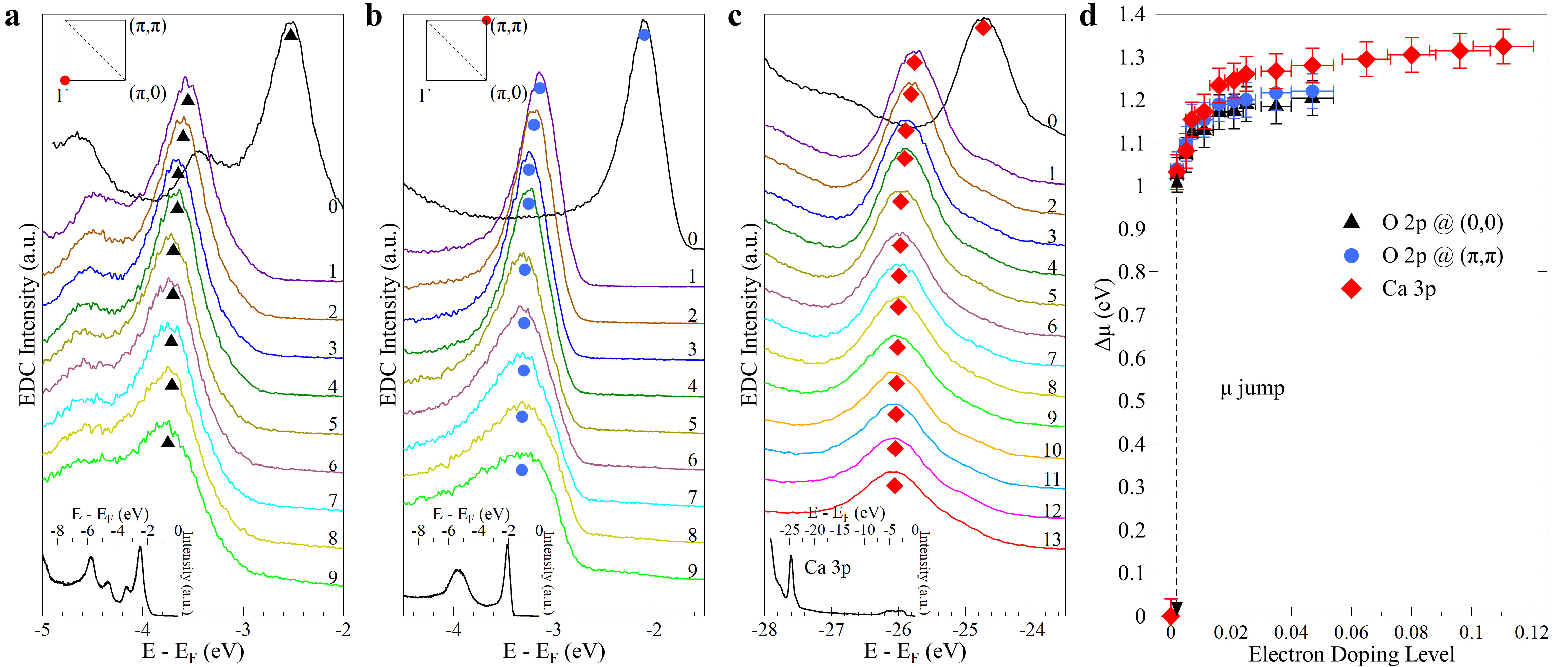}
\end{center}
\caption{{\bf Measured chemical potential shift with electron doping in Ca$_3$Cu$_2$O$_4$Cl$_2$.} (a-c) Doping evolution of O 2$p$ orbitals at $\Gamma$ (a) and ($\pi$,$\pi$) (b), and Ca 3$p$ core level (c). Symbols in (a-c) indicate the respective spectral peaks. The photoemission spectra of the undoped parent compound in a larger energy scale are shown in the respective insets. (d) Chemical potential shift with electron doping determined from the three types of measurements in (a), (b) and (c). The chemical potential shift for the undoped parent compound is set at zero as a reference. The chemical potential shifts obtained from the three cases give a similar result with a jump of $\sim$1 eV in the initial Rb deposition stage. Error bars reflect the uncertainty in determine the EDC peak positions.
}
\end{figure*}

\begin{figure*}[tbp]
\begin{center}
\includegraphics[width=1\columnwidth,angle=0]{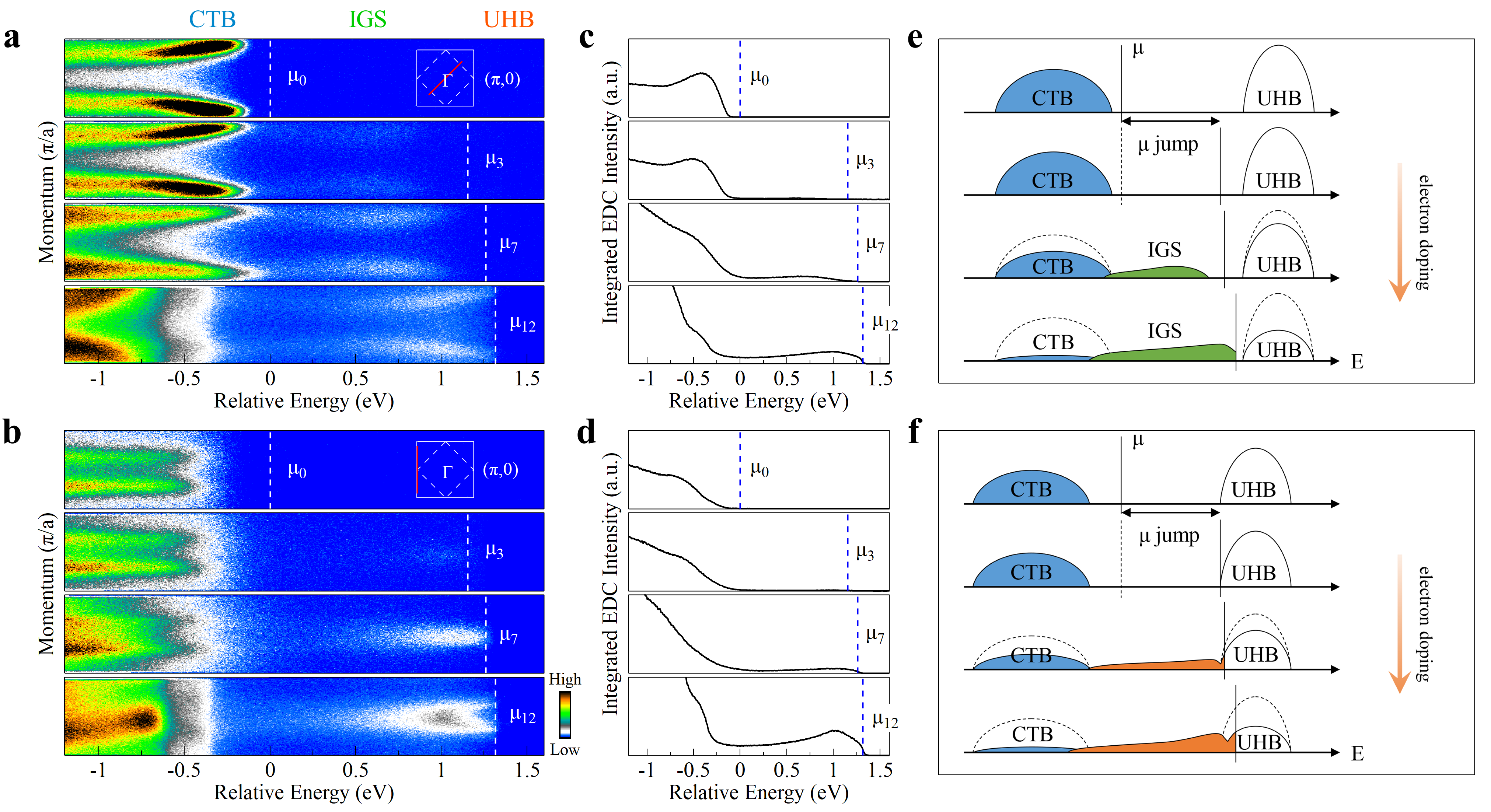}
\end{center}
\caption{{\bf Momentum-resolved electronic structure evolution picture of Ca$_3$Cu$_2$O$_4$Cl$_2$ with electron doping.}  (a) Representative band structures at different electron doping levels, corresponding to  Rb deposition stages of  0, 3, 7 and 12,  measured along the  (-$\pi$/2,-$\pi$/2)-($\pi$/2,$\pi$/2) nodal cut.  (b) Corresponding band structures at different electron doping levels measured along the (-$\pi$,-$\pi$)-(-$\pi$,$\pi$) cut.  The horizontal axes in  (a) and (b) are plotted on a relative energy scale where the chemical potential position of the original undoped CCOC326 is set at zero, and the energy position of others is referenced to the chemical potential shift $\mu$ shown in Fig. 4d.  The chemical potential positions for each doping are marked by white dashed lines. The corresponding momentum cuts for the bands are shown in the upper-right insets.  (c,d) Doping evolution of the momentum-integrated EDCs which are obtained from (a) and (b), respectively,  by integrating the spectral weight over the entire momentum range of the cuts shown as red lines in the insets of  (a,b).  (e,f) Schematic electronic structure evolution with electron doping along the  (-$\pi$/2,-$\pi$/2)-($\pi$/2,$\pi$/2) nodal direction, and along the (-$\pi$,-$\pi$)-(-$\pi$,$\pi$) zone face, respectively. The three spectral components, charge transfer band (CTB), in-gap states (IGS) and upper Hubbard band (UHB), are also labelled on top of (a) for comparison.
}
\end{figure*}

\end{document}